\documentclass[epj,referee,amsmath,amssymb]{svjour}
\usepackage{latexsym}
\usepackage{graphics}
\begin{document}
\title{Pressure dependence of the Shubnikov-de Haas oscillation spectrum of
$\beta$''-(BEDT-TTF)$_4$(NH$_4$)[Cr(C$_2$O$_4$)$_3$]$\cdot$DMF}

\author{ David Vignolles\inst{1}, Vladimir N. Laukhin\inst{2,3}, Alain
Audouard\inst{1}, Marc Nardone\inst{1}, Tatyana G.
Prokhorova\inst{4}, Eduard B. Yagubskii\inst{4} and Enric
Canadell\inst{3}
}                     
%
\mail{audouard@lncmp.org}
\institute{Laboratoire National des Champs Magn\'{e}tiques
Puls\'{e}s \thanks{UMR 5147: Unit\'{e} Mixte de Recherche CNRS -
Universit\'{e} Paul Sabatier - INSA de Toulouse}, 143 avenue de
Rangueil, 31400 Toulouse, France \and Instituci\'{o} Catalana de
Recerca i Estudis Avan\c{c}ats (ICREA), 08010 Barcelona, Spain
\and Institut de Ci\`{e}ncia de Materials de Barcelona (ICMAB -
CSIC), Campus UAB, 08193 Bellaterra, Catalunya, Spain \and
Institute of Problems of Chemical Physics, Russian Academy of
Sciences, 142432 Chernogolovka, MD, Russia}
\date{Received: \today / Revised version: date}
%
\abstract{The Shubnikov-de Haas (SdH) oscillation spectra of the
$\beta$''-(BEDT-TTF)$_4$(NH$_4$)[Cr(C$_2$O$_4$)$_3$]$\cdot$DMF
organic metal have been studied in pulsed magnetic fields of up to
either 36 T at ambient pressure or 50 T under hydrostatic
pressures of up to 1 GPa. The ambient pressure SdH oscillation
spectra can be accounted for by up to six fundamental frequencies
which points to a rather complex Fermi surface (FS). A noticeable
pressure-induced modification of the FS topology is evidenced
since the number of frequencies observed in the spectra
progressively decreases as the pressure increases. Above 0.8 GPa,
only three compensated orbits are observed, as it is the case for
several other isostructural salts of the same family at ambient
pressure. Contrary to other organic metals, of which the FS can be
regarded as a network of orbits, no frequency combinations are
observed for the studied salt, likely due to high magnetic
breakdown gap values or (and) high disorder level evidenced by
Dingle temperatures as large as $\simeq$ 7 K.
\PACS{
      {71.18.+y}{Fermi surface: calculations and measurements; effective mass, g factor} \and
      {71.20.Rv }{Polymers and organic compounds}  \and
      {72.20.My}{Galvanomagnetic and other magnetotransport effects}
     } 
} 
\authorrunning {D. Vignolles et al.}
\titlerunning{Pressure dependence of SdH oscillation spectrum of $\beta$''-(ET)$_4$(NH$_4$)[Cr(C$_2$O$_4$)$_3$]$\cdot$DMF}
\maketitle
\section{Introduction}
\label{intro}

The family of isostructural monoclinic charge-transfer salts
$\beta$''-(BEDT-TTF)$_4$(A)[M(C$_2$O$_4$)$_3$]$\cdot$Solv have
been widely studied in the past decade \cite{Cor04}. In the above
formula, BEDT-TTF stands for
bis(ethylenedithio)tetrathiafulvalene, A is a monovalent cation (A
= H$_3$O$^+$, K$^+$, NH$_4$$^+$, etc.), M is a trivalent cation (M
= Cr$^{3+}$, Fe$^{3+}$, Ga$^{3+}$, etc.) and Solv is a solvent
molecule such as benzonitrile (C$_6$H$_5$CN), dimethylformamide
(C$_3$H$_7$NO), nitrobenzene (C$_6$H$_5$NO$_2$) and pyridine
(C$_5$H$_5$N), labelled hereafter BN, DMF, NB and P, respectively.
In the following, the compounds belonging to this family are
referred to as A-M$\cdot$Solv. The interest in this family of
compounds has been motivated by the observation of
superconductivity at ambient pressure in the H$_3$O-Fe$\cdot$BN
salt (T$_c$ = 8.5 K) \cite{Gr95}. Later on, other superconducting
salts with magnetic ions were reported for this family
\cite{Ma97,Ra01}. Besides, a metallic and ferromagnetic ground
state was achieved in the (BEDT-TTF)$_3$[MnCr(C$_2$O$_4$)$_3$]
compound \cite{Co00}. Contrary to the orthorhombic compounds with
the same generic formula and $\beta$'' packing, which are
semiconductors \cite{Ku95,Ma01}, all the monoclinic salts of this
family exhibit metallic conductivity around room temperature.
Nevertheless, a large variety of temperature-dependent behaviours
and various ground states are observed which might be connected to
details of their electronic structure.

\begin{figure} [h]                                                   
\centering \resizebox{\columnwidth}{!}{\includegraphics*{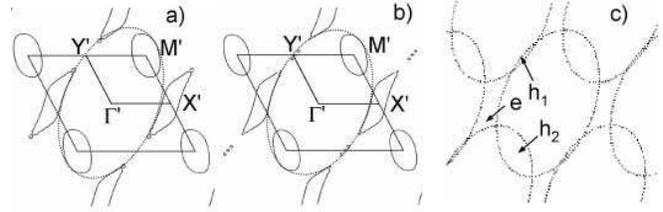}}
\caption{\label{SF} Fermi surface (FS) of (a) NH$_4$-Fe$\cdot$DMF
and (b) (NH$_4$)$_{0.75}$K$_{0.25}$-Cr$\cdot$DMF according to band
structure calculations \cite{Pr03} in which the FS is considered
on the basis of a unit cell with vectors \emph{a'} = \emph{a} and
\emph{b'} = (\emph{a} + \emph{b})/2. (c) schematic representation
of intersecting elliptic hole tubes leading to three compensated
electron ($e$) and hole ($h_1$ and $h_2$) orbits. The area of the
ellipses in dotted lines ($\bigodot$ orbits, see text) is equal to
that of the First Brillouin zone.}
\end{figure}

According to band structure calculations \cite{Pr03}, the Fermi
surface (FS) of  NH$_4$-Fe$\cdot$DMF and
(NH$_4$)$_{0.75}$K$_{0.25}$-Cr$\cdot$DMF salts originates from
quasi two-dimensional (2D) hole elliptic orbits, labelled
$\bigodot$ in the following, whose cross section is equal to the
first Brillouin zone (FBZ) area (see Fig. \ref{SF}). In the case
of NH$_4$-Fe$\cdot$DMF, these orbits intersect along the
$(a'^*+b'^*)$ direction \footnote{In Ref. \cite{Pr03}, the FS is
considered on the basis of a unit cell with vectors \emph{a'} =
\emph{a}, \emph{b'} = (\emph{a} + \emph{b})/2 and \emph{c'} =
\emph{c}. This unit cell contains four BEDT-TTF molecules.},
leading to one electron and one hole compensated orbit with a
cross section area of 8.8 percent of the FBZ one. Analogous FS
topology has also been reported for the superconducting
H$_3$O-Fe$\cdot$BN salt \cite{Ku95}. These calculations are in
agreement with the Shubnikov-de Haas (SdH) oscillation spectrum of
the H$_3$O-Ga$\cdot$NB salt for which only one frequency was
reported \cite{Ba04}. Nevertheless, as pointed out in Ref.
\cite{Pr03}, the $\bigodot$ orbits may also intersect in the
$b'^*$ direction leading to one or more additional orbits around
the Y' point of the FBZ [see Figs. \ref{SF}(b) and (c)]. This
picture holds for the NH$_4$-Fe$\cdot$DMF salt, for which the SdH
oscillation spectra can be interpreted on the basis of three
compensated orbits with cross section areas of 1.2, 4.8 and 6
percent of the FBZ area that are therefore connected by a linear
combination \cite{Au04}. However, the FS of other compounds of
this family may be more complicated since four frequencies
corresponding to orbit's area in the range 1.1 to 8.5 percent of
the FBZ area were reported for the H$_3$O-M$\cdot$P (M = Cr, Ga,
Fe) salts \cite{Co04}. In this latter case, a density wave ground
state, responsible for the observed strongly non-monotonous
temperature dependence of the resistance, has been invoked in
order to account for this discrepancy. However, only two
frequencies were observed for the H$_3$O-M$\cdot$NB (M = Cr, Ga)
salts \cite{Ba05}. Additional combination frequencies, typical of
coupled 2D orbits networks, linked to the field-induced chemical
potential oscillation \cite{mu} and (or) field-dependent Landau
level broadening \cite{LLB} were also reported \cite{Au04,Co04}.

An important feature of the oscillatory spectra of most of these
compounds is the strong field-damping factor. Indeed, Dingle
temperature values (T$_D$) in the range 2 K  to 4 K were reported
for e.g. H$_3$O-M$\cdot$P \cite{Co04} which is the signature of a
significant disorder. This feature is in line with structural data
\cite{Ak02,Tu99} which indicate that terminal ethylene groups of
some of the BEDT-TTF molecules exhibit a large solvent-dependent
positional disordering. As for the compounds with the DMF solvent,
the DMF molecules themselves are also disordered \cite{Pr03}. As a
matter of fact, even larger Dingle temperatures (T$_D$ $\approx$ 4
K to 6 K) were reported for the NH$_4$-Fe$\cdot$DMF salt
\cite{Au04}.

In order to get some insight in possible connection between the FS
topology and the ground state, we report on the pressure
dependence of the SdH spectrum of the NH$_4$-Cr$\cdot$DMF salt
which, contrary to the above mentioned NH$_4$-Fe$\cdot$DMF salt
exhibits a metallic conductivity down to about 10 K. Even though
the overall behaviour of the resistivity as the temperature varies
is unaffected by applied pressures up to 1 GPa, which suggests
that the ground state remains unchanged in this pressure range, it
is shown that the FS topology is very sensitive to applied
pressure.

\section{Experimental}
The crystals studied, labelled  $\#$ 1 to $\# $ 3 in the
following, were elongated hexagonal platelets with approximate
dimensions (0.6 $\times$ 0.4 $\times$ 0.25)~mm$^3$ for crystal
$\#$ 2 and (0.4 $\times$ 0.2 $\times$ 0.1)~mm$^3$ for crystals
$\#$ 1 and $\#$ 3, the largest faces being parallel to the
conducting \emph{ab}-plane. Magnetoresistance experiments were
performed in pulsed magnetic field of up to 36 T for crystals $\#$
1 and $\#$ 2 and 50 T for crystal $\#$ 3, with pulse decay
duration of 0.78 s and 0.32 s, respectively, in the temperature
range from 2 K to 4.2 K. For crystals $\#$ 1 and $\#$ 3, the
magnetic field was applied normal to the conducting plane whereas
a sample holder rotating about an axis perpendicular to the
magnetic field allowed a change of the direction of the magnetic
field with respect to the conducting plane for crystal $\#$ 2.
Crystal $\#$ 3 was studied under hydrostatic pressure in an anvil
cell designed for isothermal measurements in pulsed magnetic
fields \cite{Na01}. The pressure applied at room temperature was
estimated from the primary pressure value calibrated beforehand
with a manganin piezoresistive sensor. The maximum pressure
reached in the experiments was 1 GPa at low temperature, taking
into account a pressure decrease of 0.1 GPa on cooling. Electrical
contacts to the crystal were made using annealed platinum wires of
20 $\mu$m in diameter glued with graphite paste. Alternating
current (1 to 17 $\mu$A, 20 kHz) was injected parallel to the
\emph{c}* direction (interlayer configuration). A lock-in
amplifier with a time constant of 100 $\mu$s was used to detect
the signal across the potential contacts.

Analysis of the oscillatory magnetoresistance is based on discrete
Fourier transforms and direct fittings of the magnetoresistance
data. Discrete Fourier transforms are calculated with a Blackman
window in a given field range from $B_{min}$ to $B_{max}$. The
absolute value of the amplitude ($A_i$) of the Fourier component
with frequency $F_i$ is determined, for a mean field value
$B_{mean}=2/(1/B_{min}+1/B_{max})$, from the amplitude of the
discrete Fourier transform ($A_{calc}$) as $A_i =
4A_{calc}/0.84(1/B_{min}-1/B_{max})$. The direct fitting method
was the following: a low order polynomial (typically 2th order)
together with one Fourier component with initial frequency close
to one of the frequencies detected in experimental data is
introduced. In the case where residuals still exhibit oscillatory
features, a subsequent oscillation is introduced with a frequency
close to one of the main frequencies detected in the residuals.
The procedure is repeated until either not any oscillatory
component can be detected in the residuals or the oscillatory
component corresponds to high order harmonics of already detected
frequencies. Finally, the order of the polynomial, which account
for the background, is increased up to at most the 4th order. It
has been checked that neither the detected frequencies depend
appreciably on the order in which the various components are
introduced in the fittings nor the magnitude of the background
significantly depends on the order of the polynomial.

\begin{figure}                                                       
\centering \resizebox{\columnwidth}{!}{\includegraphics*{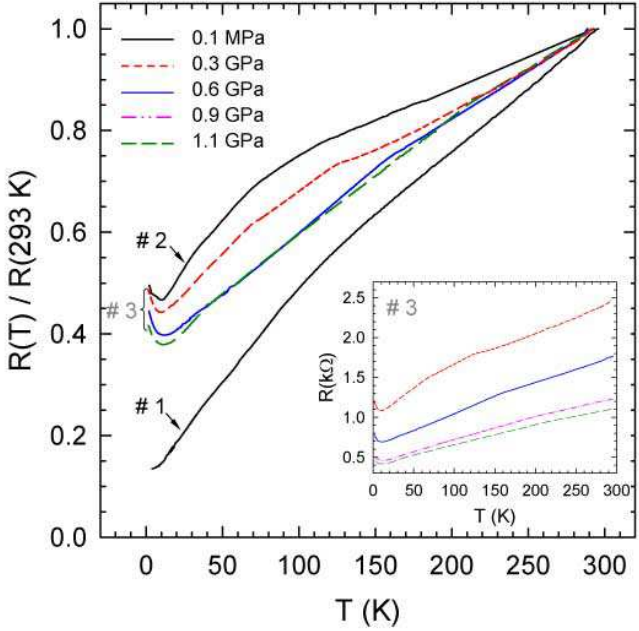}}
\caption{Temperature dependence of the zero-field resistance
normalized to the room temperature value for various pressures
measured at room temperature. The crystal number (see text) is
indicated on the curves. The inset displays the temperature
dependence of the crystal $\#$ 3 resistance for the same
pressures.} \label{RT}
\end{figure}

\section{Results and discussion}
\label{sec:results}

\begin{figure} 
\centering
\resizebox{\columnwidth}{!}{\includegraphics*{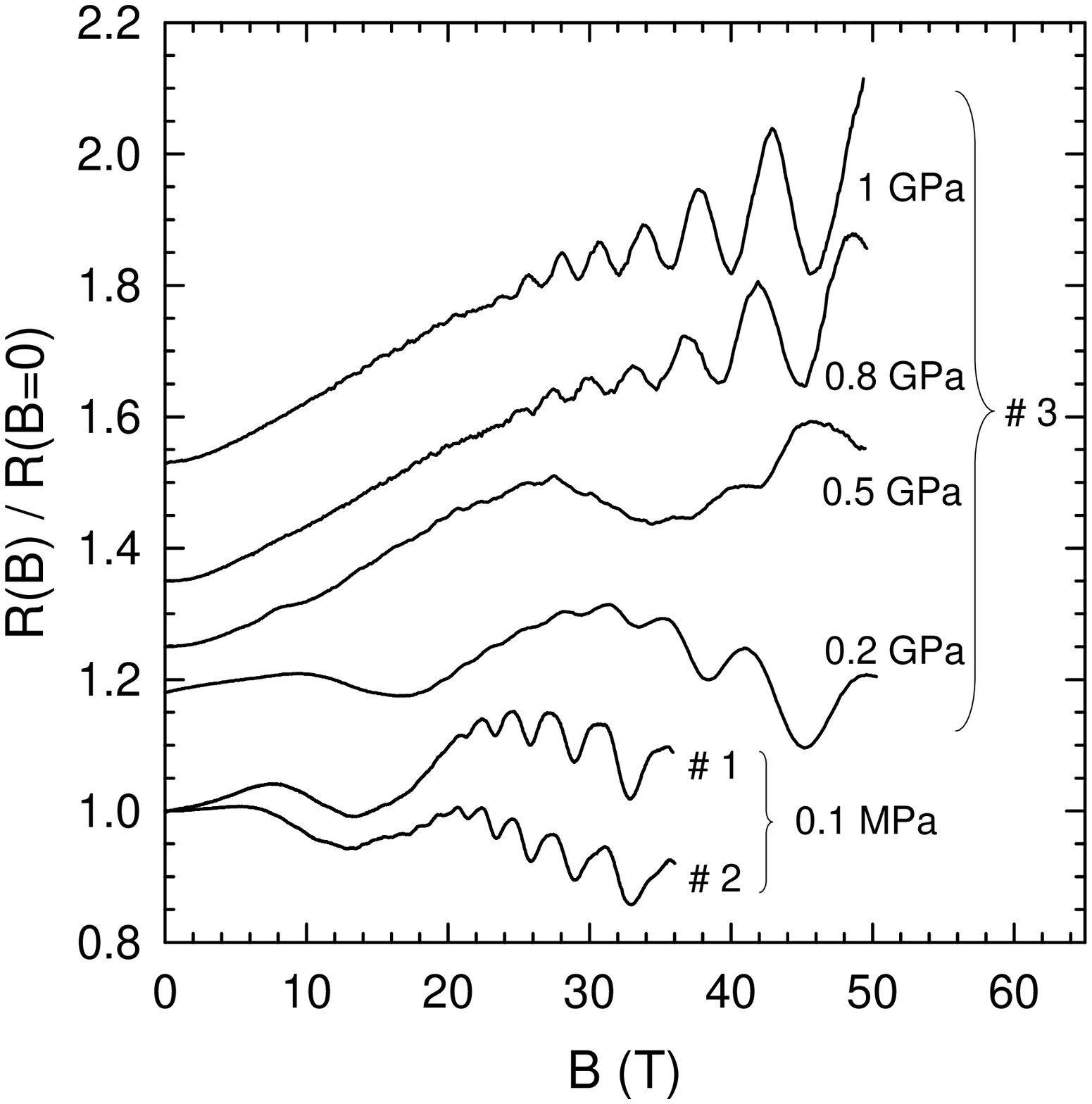}}
\caption{\label{R(B)_P} Magnetoresistance at 2 K for various
pressures (the low temperature values of which are given). The
crystal number (see text) is indicated. Curves have been shifted
from each other for clarity.}
\end{figure}

The zero-field temperature dependence of the resistance of the
studied crystals is displayed in Figure \ref{RT}. A metallic
behaviour is observed down to about 10 K in the pressure range
explored. The crystal-dependent resistance ratio R(10 K)/R(293 K)
is in the range 0.15 $\div$ 0.5. This behaviour, which is in
agreement with the data of \cite{Pr03}, is at variance with the
strongly non-monotonic temperature dependence reported for other
salts such as H$_3$O-M$\cdot$Solv (M = Ga, Cr; Solv = NB, P)
\cite{Ba04,Ak02}, NH$_4$-Fe$\cdot$DMF \cite{Au04} or
H$_3$O-M$\cdot$P (M = Ga, Fe) \cite{Co04}. In the latter case, a
metal-density wave transition has been suggested to occur around
150 K in order to account for the observed behaviour. Below 10 K,
a small resistance rise is observed that might be linked to
disorder. Although a significant pressure dependence of the
resistance is observed (see the inset of Figure \ref{RT}) in the
whole temperature range explored [e.g. dln(R)/dP $\simeq$ 1
GPa$^{-1}$ at room temperature], the applied pressure has only a
minor effect on the resistance ratio. Magnetoresistance data at 2
K are presented in Figure \ref{R(B)_P} for various pressures. In
addition to magnetoresistance oscillations, a non-monotonous
behaviour which appears as a slow undulation is observed up to 0.6
GPa.

    In the following, we concentrate first on the ambient
pressure SdH oscillation spectra (see Section \ref{sec:Ambient
pressure magnetoresistance}). The pressure dependence of the
magnetoresistance oscillation spectra is considered in Section
\ref{sec:Pressure-dependent magnetoresistance}.

\subsection{Ambient pressure oscillatory spectrum}
\label{sec:Ambient pressure magnetoresistance}

\begin{figure}                                                     
\centering
\resizebox{\columnwidth}{!}{\includegraphics*{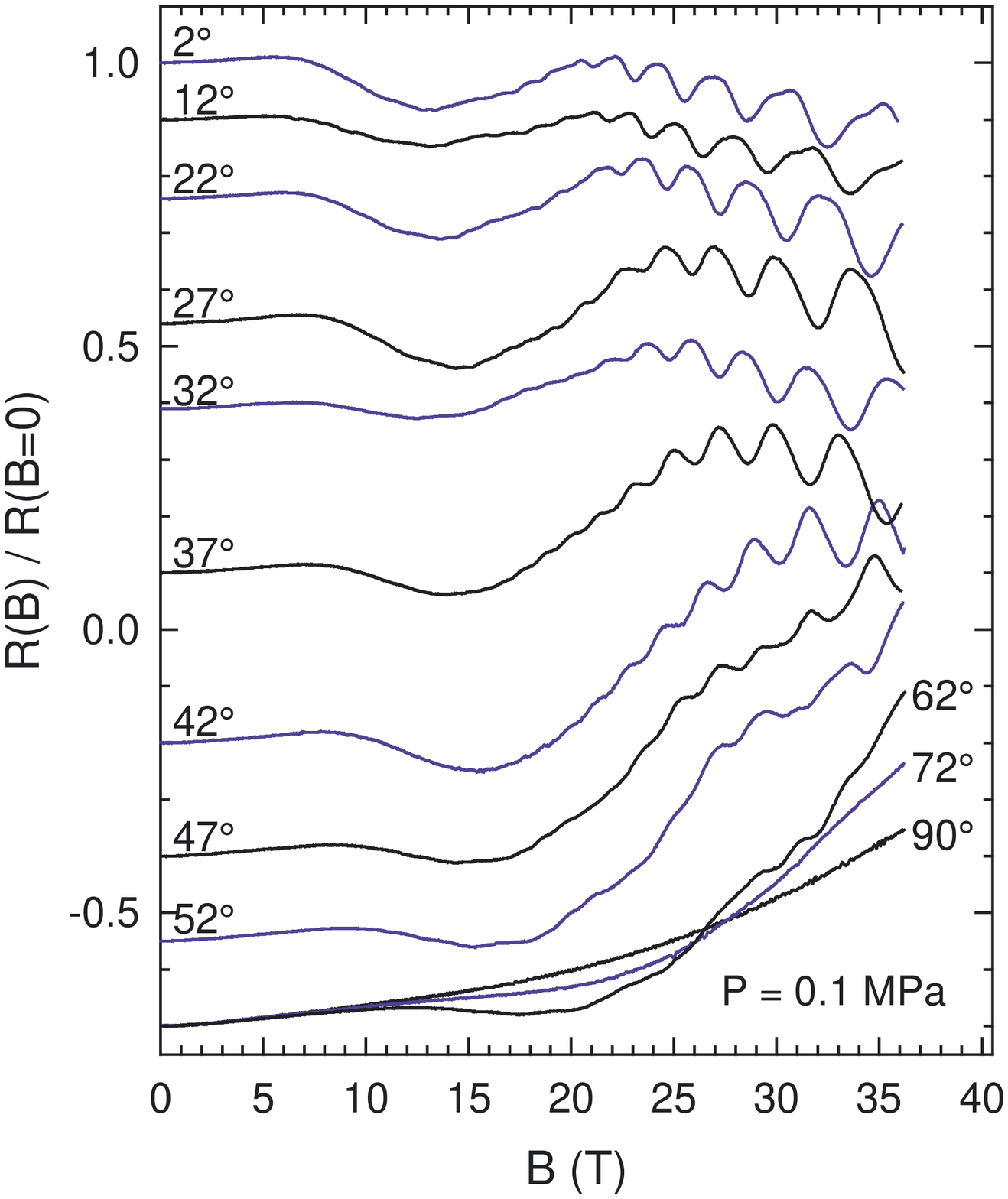}}
\caption{\label{R(B)-Pamb_angle} Ambient pressure
magnetoresistance of crystal $\#$ 2 at 2 K for various directions
of the magnetic field. The angle between the magnetic field and
the conducting plane is indicated. Curves have been shifted down
from each other for clarity.}
\end{figure}

\begin{figure*}                                                     
\centering
{\includegraphics*{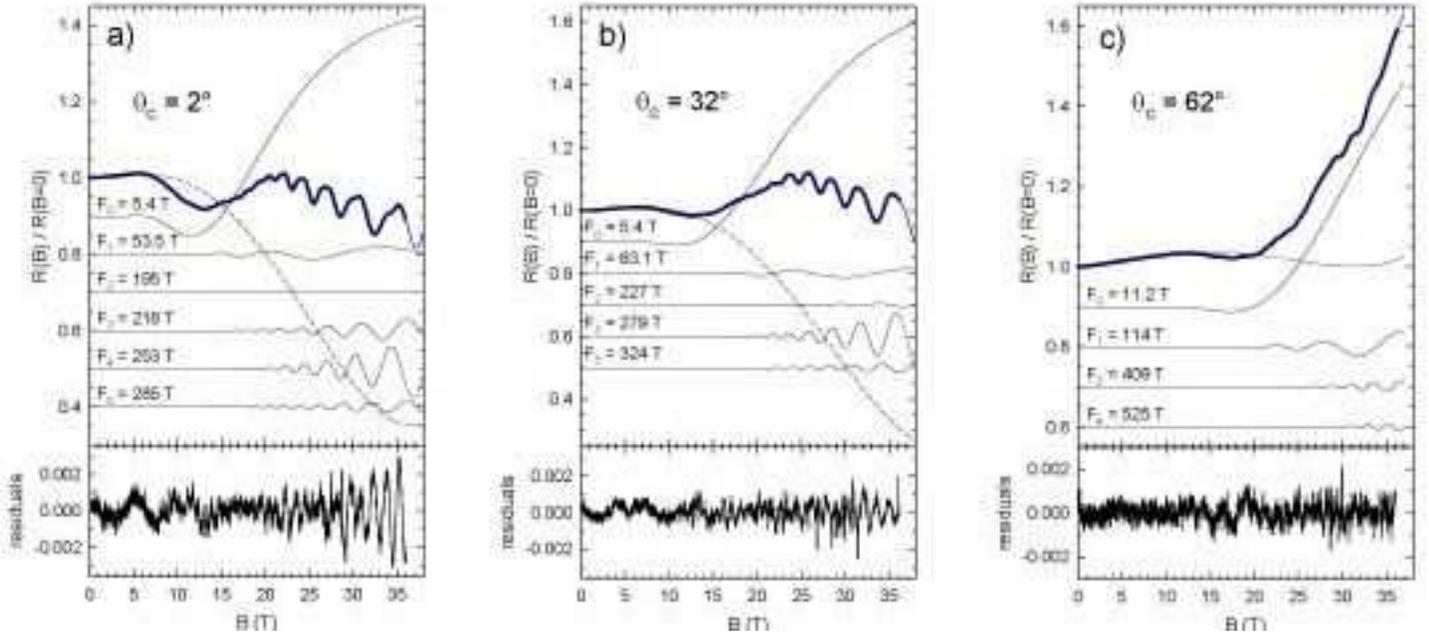}} \caption{\label{fit_m}
Magnetoresistance data of crystal $\#$ 2 at ambient pressure
(thick black solid lines). Thin blue solid lines are corresponding
best fits of Eq. \ref{approxLK}. Thin black dashed and solid lines
displays the background magnetoresistance (4$^{th}$ order
polynomial) and the contribution of each frequency (index $i$ in
Eq. \ref{approxLK}), respectively, entering the fits. Curves have
been shifted down from each other for clarity. Data are collected
at 2 K for various directions of the magnetic field. Residuals
values are given by [R(B) / R(B = 0)]$_{experimental}$ - [R(B) /
R(B = 0)]$_{fit}$.}
\end{figure*}

\begin{figure}                                                       
\centering
\resizebox{\columnwidth}{!}{\includegraphics*{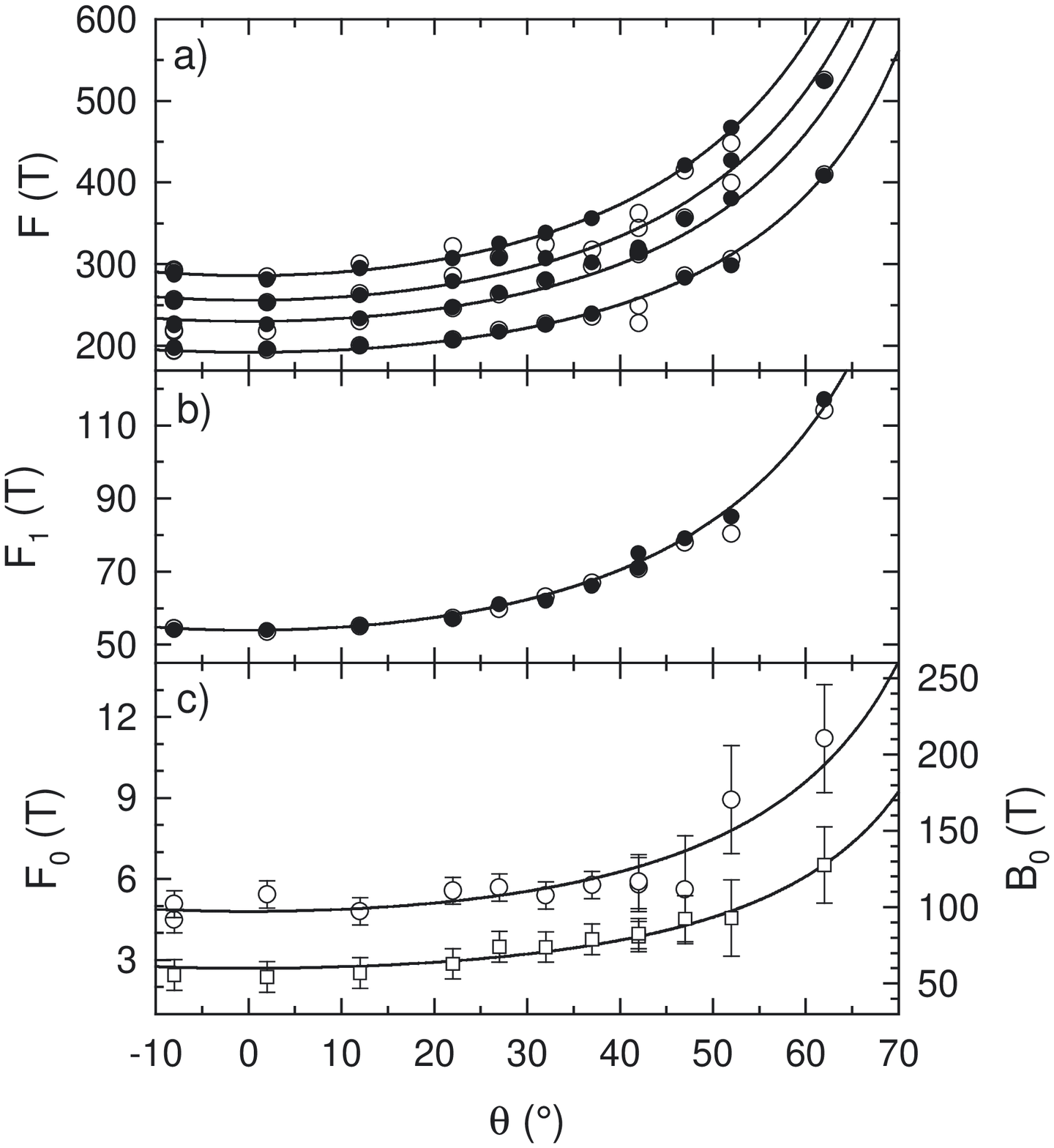}}
\caption{\label{F(a)} Angle dependence of the frequencies observed
at ambient pressure. (a) and (b) displays F$_2$ to  F$_5$ and
F$_1$, respectively. Open and closed symbols are deduced from best
fits of Eq. \ref{approxLK} (see Fig. \ref{fit_m}) and Fourier
analysis, respectively. (c) displays the angle variation of F$_0$
(circles) and of the parameter B$_0$ (squares) deduced from best
fits of Eq. \ref{approxLK} to the data. }
\end{figure}

Figure \ref{R(B)-Pamb_angle} displays the normalized
magnetoresistance of sample $\#$ 2 at a temperature of 2 K for
various directions of the magnetic field. The nature of the above
mentioned slow undulation, whose extremes are angle-dependent,
needs clarification since it can be due to either some
non-monotonic behaviour of the background magnetoresistance or to
a slow SdH oscillation linked to the presence of a very small
orbit. In any case, it makes the determination of the background
magnetoresistance difficult. This background should be properly
removed from the magnetoresistance data in order to avoid a large
zero-frequency peak liable to hamper the extraction of reliable
oscillatory data at low frequency (say below few tens of teslas)
by Fourier analysis. Moreover, as developed later on and in
agreement with data from other compounds of the same family
\cite{Au04,Co04}, large Dingle temperatures are observed. This
feature leads to a steep field dependence of the oscillation
amplitude which reduces the field range in which oscillations can
be detected and therefore broadens the various components' peaks
appearing in the Fourier transforms. For these reasons, in
addition to Fourier analysis, information on the oscillatory
spectra were extracted from direct fittings of the
Lifshits-Kosevich formula (LK) to the magnetoresistance data,
assuming the background magnetoresistance (R$_{bg}$) can be
approximated by a 4$^{th}$ order polynomial.

According to the LK formula, the oscillatory magnetoresistance of
a metal whose FS is composed of several 2D orbits is given by:

\begin{eqnarray}
\label{LK} \frac{R(B)}{R_{bg}} = 1 +
\sum_{i}a_i\times\nonumber\\\sum_{\lambda=1}^{\infty}R_{Ti\lambda}R_{Di\lambda}R_{MBi\lambda}R_{Si\lambda}
(-1)^{\lambda+1}cos[2{\pi}\lambda(\frac{F_i}{B}-\gamma_i)]
\end{eqnarray}

where $\lambda$ is the harmonic order. F$_i$ and $\gamma_i$ are
the frequency and the phase factor of the oscillation linked to
the orbit $i$. Equation \ref{LK} assumes that the oscillations
amplitude is small so that R(B)/R$_{bg}$ - 1 $\simeq$ 1 -
$\sigma$(B)/$\sigma_{bg}$ and that the Hall effect is either
negligible (which is actually the case for interlayer
magnetoresistance measurements with the magnetic field normal to
the conducting plane) or contributes to the background
magnetoresistance, only. The thermal (for a 2D FS), Dingle,
magnetic breakthrough (MB) and spin (S) damping factors are
respectively given by \cite{Sh84}:

\begin{eqnarray}
\label{RT} R_{Ti\lambda} = \frac{{\alpha}T{\lambda}m{_i^*}(\theta
= 0)}{Bcos{\theta}sinh[{\alpha}T{\lambda}m{_i^*}(\theta = 0)/B
cos{\theta}]} \\\label{RD}R_{Di} =
exp[-{\alpha}T_D{\lambda}m{_i^*}(\theta = 0)/B cos{\theta}]\\
\label{RMB}R_{MBi{\lambda}} =
exp(-\frac{t_iB_{MB}}{2Bcos{\theta}})[1-exp(-\frac{B_{MB}}{Bcos{\theta}})]^{b_i/2}\\
\label{RS}R_{Si{\lambda}} =
\mid\cos({\pi}{\lambda}\mu/\cos\theta)\mid
\end{eqnarray}

where $\alpha$ = 2$\pi^2$m$_e$k$_B$/e$\hbar$ ($\simeq$ 14.69 T/K),
m$_i^*$ is the effective mass normalized to the free electron mass
m$_e$, $\theta$ is the angle between the field direction and the
conducting plane, T$_D$ is the Dingle temperature, $\mu$ =
g$^*$m${_i^*}$($\theta$ = 0)/2, g$^*$ is the effective Land\'{e}
factor and B$_{MB}$ is the MB field. Integers $t_i$ and $b_i$ are
respectively the number of tunnelling and Bragg reflections
encountered along the path of the quasiparticle. In the high
$Tm{_i^*}/B$ range, for $R_{bg}$ close to $R(B=0)$ and assuming
large (low) $B_{MB}$  values and $t_i$ = 0 ($b_i$ = 0), in which
case $R_{MBi}$ = 1, Equation \ref{LK} can be approximated as:

\begin{eqnarray}
\label{approxLK}R(B) \simeq R_{bg}
+\sum_{i}{\frac{A_i}{B}}\times\nonumber\\
\sum_{{\lambda}=1}^{\infty}exp(-{\lambda}\frac{B_i}{B})cos[2{\pi}p(\frac{F_i}{B}-\gamma_i)]
\end{eqnarray}

where $A_i$ is a field-independent parameter, including in
particular the contribution of the spin damping factor, and $B_i$
= $\alpha(T+T_D)m_{i}^*/cos{\theta}$. It should be kept in mind
that, in addition to the above mentioned approximations, some
deviations of the magnetoresistance oscillations from the LK
formula are observed for 2D FS's, in particular for clean crystals
at low T/B values. As a consequence, Equation \ref{approxLK} may
not yield reliable values of the $A_i$ and $B_i$ parameters.
Nevertheless, as reported hereafter, this equation is useful in
order to identify the various components entering the Fourier
spectra with a restricted number of free parameters.

\begin{figure} 
\centering
\resizebox{\columnwidth}{!}{\includegraphics*{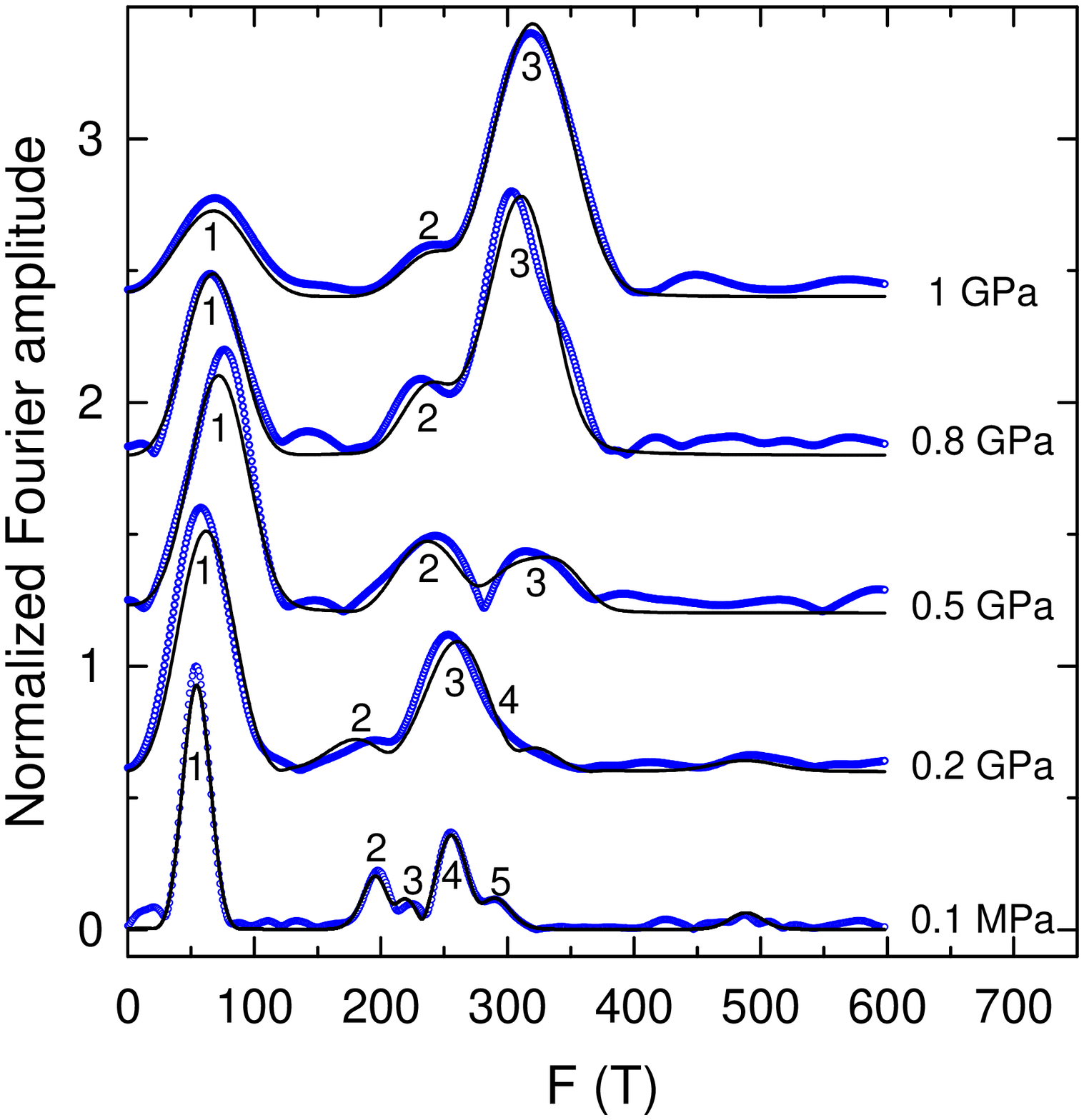}}
\caption{\label{TF_P} Fourier spectra of the oscillatory
magnetoresistance of crystals $\#$ 2 (at ambient pressure) and
$\#$ 3 (under applied pressure) at 2 K (blue symbols) and of the
corresponding fits of Eq. \ref{approxLK} (solid black lines). The
magnetic field range is 8 T - 36 T for the data at 0.1 MPa, 15 T -
50 T for the data in the range 0.2 GPa to 0.8 GPa and 18 T - 50 T
for the data at 1 GPa. The spectra, which are shifted from each
other for clarity, have been normalized to the component with the
highest amplitude. The labels correspond to the frequency index
used in the text.  }
\end{figure}

Examples of best fits of Equation \ref{approxLK} to
magnetoresistance data recorded at ambient pressure are displayed
in Figure \ref{fit_m}. According to these data, six frequencies
labelled F$_0$ to F$_5$ in the following, enter the oscillatory
part of the magnetoresistance for magnetic field direction not too
far from the normal to the conducting plane (up to $\theta$
$\simeq$ 30$^{\circ}$), the lower frequencies F$_0$ to F$_2$ being
perceptible up to $\theta \simeq$ 65$^{\circ}$. The residuals
displayed in the bottom part of the figures either only contain
high order harmonics or do not reveal any periodic component. E.g.
the high field part of the residuals of data in Figure
\ref{fit_m}(a) is dominated by the 2$^{nd}$ harmonic of F$_3$ and
the 3$^{rd}$ harmonic of F$_4$.  As reported in Figure \ref{F(a)},
the deduced frequencies follow the orbital behaviour expected for
a 2D FS. Remarkably, the slow undulation is accounted for by an
SdH oscillation with the frequency F$_0$. In addition, a clear
orbital behaviour of the parameter $B_0$ is observed in Figure
\ref{F(a)}(c), as it is the case for the B$_i$ parameters relevant
to the other components. A strongly negative background
magnetoresistance is deduced from the fits for low $\theta$ values
[see Figs. \ref{fit_m}(a) and (b)]. Although a strongly negative
magnetoresistance has already been reported above the critical
field in the H$_3$O-M$\cdot$NB superconducting salts (M = Ga, Cr)
\cite{Ba04}, it should be noticed that the main part of the
magnetoresistance results from the contributions of F$_0$ and the
background which, according to the data in Figure \ref{fit_m},
have opposite variation above $\sim$ 15 T. In addition, the
quantum limit is reached at a few tens of teslas for this
frequency. In such a case, a significant error on their amplitude
cannot be excluded. The frequency values deduced from the fits of
the magnetoresistance data for crystals $\#$ 1 and $\#$ 2 are
displayed in Table \ref{table}. Since the frequency F$_0$ is very
low, Fourier analyses have been performed subtracting the
contributions of both the background magnetoresistance $R_{bg}$
and the oscillation with frequency F$_0$ from the
magnetoresistance data. An example is given in Figure \ref{TF_P}:
a good agreement between Fourier analysis of the fits of Equation
\ref{approxLK} and of the experimental data is observed.

\begin{table}                                                            
\caption{\label{table}Frequencies, reduced to $\theta$ =
0$^{\circ}$, deduced from the best fits of Eq. \ref{approxLK} to
magnetoresistance data at ambient pressure and from Fourier
analysis (see Fig. \ref{TF_P}). $i$ is the frequency index
appearing in Eqs. \ref{LK} and \ref{approxLK}.}
\begin{tabular}{ccccc}
\hline
&\multicolumn{2}{c} {crystal $\#$ 1}&\multicolumn{2}{c} {crystal $\#$ 2} \\
i &fit&FT&fit&FT\\
\hline
0&6.2$\pm$1.0&&5.0$\pm$0.5\\
1&54$\pm$1&54.8$\pm$0.5&53.5$\pm$1.5&53.7$\pm$1.0\\
2&202$\pm$3&200$\pm$2&193$\pm$3&196$\pm$3\\
3&229$\pm$6&230$\pm$5&218$\pm$7&226$\pm$4\\
4&257$\pm$5&250$\pm$2&253$\pm$5&253$\pm$4\\
5&295$\pm$3&287$\pm$5&288$\pm$6&284$\pm$6\\
\hline
\end{tabular}
\end{table}

F$_0$ may correspond to an orbit with a very small cross section
amounting to 0.1 percent of the FBZ area only. Such a low value is
compatible with band structure calculations since the cross
section values of the orbits close to the Y' point of the FBZ in
Figure \ref{SF}(b) are of the same order of magnitude or even
smaller. The values of the other observed frequencies (F$_1$ to
F$_5$) correspond to orbital areas in the range 1 to 7 percent of
the FBZ area. Such values are of the same order of magnitude as
those deduced from the data of Refs. \cite{Au04,Co04}. For
example, F$_1$ is very close to the frequencies F$_a$ = 48 T and
F$_{\alpha}$ = 38 to 50 T reported for the NH$_4$-Fe$\cdot$DMF
\cite{Au04} and H$_3$O-M$\cdot$P \cite{Co04} salts, respectively,
while F$_4$ is very close to the frequency F$_b$ = 248 T reported
for NH$_4$-Fe$\cdot$DMF \cite{Au04}. It can be deduced from the
data in Table \ref{table} that F$_0$+F$_1$+F$_2$+F$_4$ is equal to
F$_3$+F$_5$ within the error bars, as expected for a compensated
metal. However, even in the case where this latter relationship is
not fortuitous, the large number of observed frequencies cannot be
fully understood on the basis of the band structure calculations
displayed in Figure \ref{SF}. Nevertheless, a FS based on the
intersection of elliptic 2D tubes scheme might still account for a
large number of orbits. If true, the actual picture would be less
naive than that displayed in Figure \ref{SF}(c) and, in any case,
this point needs a more detailed determination of the FS topology.
Other relationships such as F$_4$ = F$_1$ + F$_2$ or F$_5$ = F$_1$
+ F$_4$ are also observed (see Table \ref{table}). Still in the
case where they are not fortuitous, such linear combinations could
indicate that frequencies F$_4$ and (or) F$_5$ are linked to
either MB orbits or frequency combinations. However, since band
structure calculations cannot yield detailed FS topology, no
reliable conclusion can be drawn regarding the presence of MB
orbits. Effective masses deduced from the temperature dependence
of the Fourier component's amplitudes are in the range from 0.4 to
1.2 free electron mass (see Table \ref{table2}) which is of the
same order of magnitude as for other salts of this family
\cite{Au04,Co04}. Dingle temperature values are high (T$_D$ $\sim$
7 K) which certainly rules out frequency combinations due to an
oscillation of the chemical potential \cite{mu}. Otherwise,
although crystals $\#$ 1 and $\#$ 2 exhibit significantly
different residual resistance ratios, their Dingle temperature are
rather close. This feature is in line with the statement of Ref.
\cite{Co04} that the crystals are composed of a mixture of
insulating and metallic domains, although a metallic conductivity
is observed in the present case. Within this picture, the
temperature dependence of the resistance reflects the relative
parts of metallic and insulating domains while the oscillatory
behaviour is only related to the metallic parts which are in turn
characterized by a large disorder as indicated by the large
measured Dingle temperatures.

\subsection{\label{sec:Pressure-dependent magnetoresistance}Pressure-dependent oscillatory spectra}

\begin{figure*}                                                    
\centering
{\includegraphics*{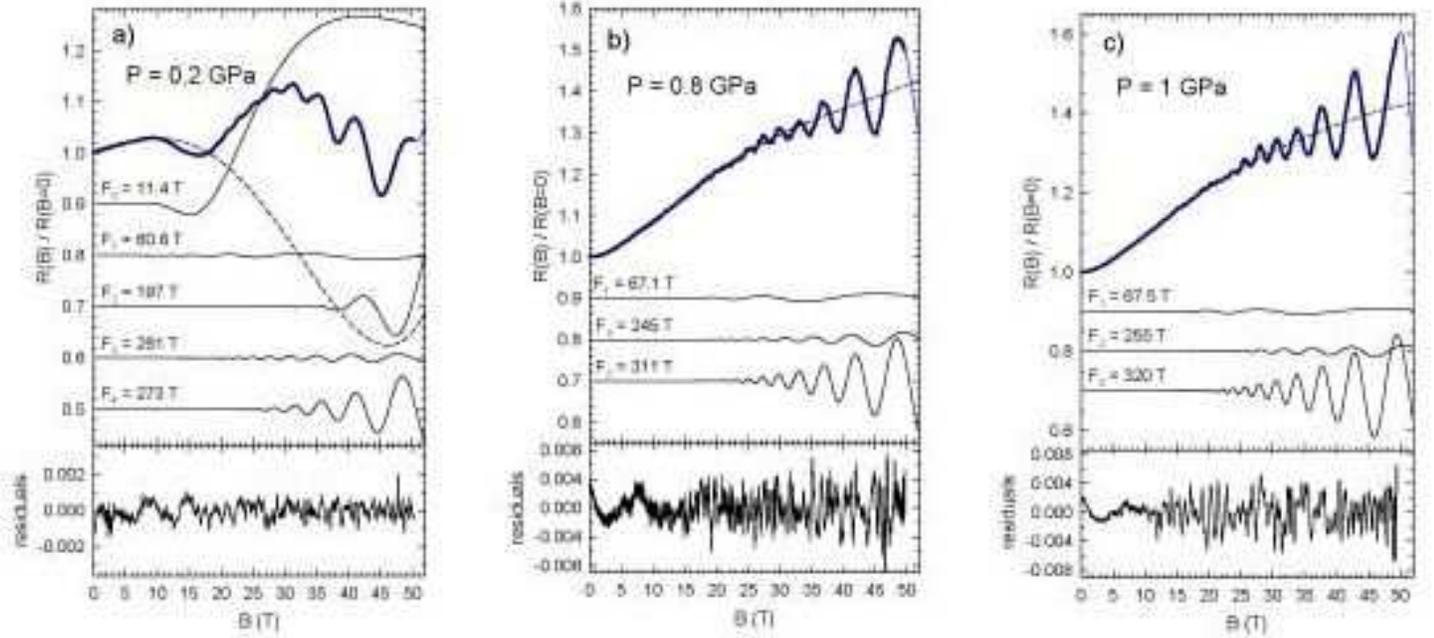}}
\caption{\label{fit_m_appliedP} Same as Fig. \ref{fit_m} for
crystal $\#$ 3 at various applied pressures.}
\end{figure*}

\begin{table*}                                                            
\caption{\label{table2}Effective masses and Dingle temperatures
deduced from temperature and field dependence, respectively, of
the amplitude of the various oscillations observed. $i$ is the
frequency index appearing in Eqs. \ref{LK} and \ref{approxLK}. The
crystal numbers are indicated.}
\begin{tabular}{ccccccccccc}
\hline
&\multicolumn{2}{c} {0.1 MPa ($\#$ 1)}&\multicolumn{2}{c} {0.1 MPa ($\#$ 2)}&\multicolumn{2}{c} {0.5 GPa ($\#$ 3)}&\multicolumn{2}{c} {0.8 GPa ($\#$ 3)}&\multicolumn{2}{c} {1 GPa ($\#$ 3)}\\
i &m$^*_i$&T$_D$&m$^*_i$&T$_D$&m$^*_i$&T$_D$&m$^*_i$&T$_D$&m$^*_i$&T$_D$\\
\hline
1&0.4$\pm$0.2&7.5$\pm$5.0&0.4$\pm$0.2&6.5$\pm$4.0&0.5$\pm$0.05&8$\pm$2&0.4$\pm$0.2&5.0$\pm$2.5&0.60$\pm$0.15&4$\pm$2\\
2&0.4$\pm$0.1&6$\pm$4&0.4$\pm$0.1&7$\pm$4.5&0.6$\pm$0.2&6.5$\pm$3.5&0.80$\pm$0.25&7.5$\pm$4.0&0.8$\pm$0.2&\\
3&0.7$\pm$0.2&&0.6$\pm$0.2&&0.9$\pm$0.2&&1.10$\pm$0.25&8.5$\pm$3.5&0.8$\pm$0.2&9$\pm$3\\
4&1.2$\pm$0.2&6$\pm$3&1.2$\pm$0.2&6$\pm$2&\\
5&0.65$\pm$0.20&&0.65$\pm$0.20&\\
\hline
\end{tabular}
\end{table*}

\begin{figure}                                                      
\centering
\resizebox{\columnwidth}{!}{\includegraphics*{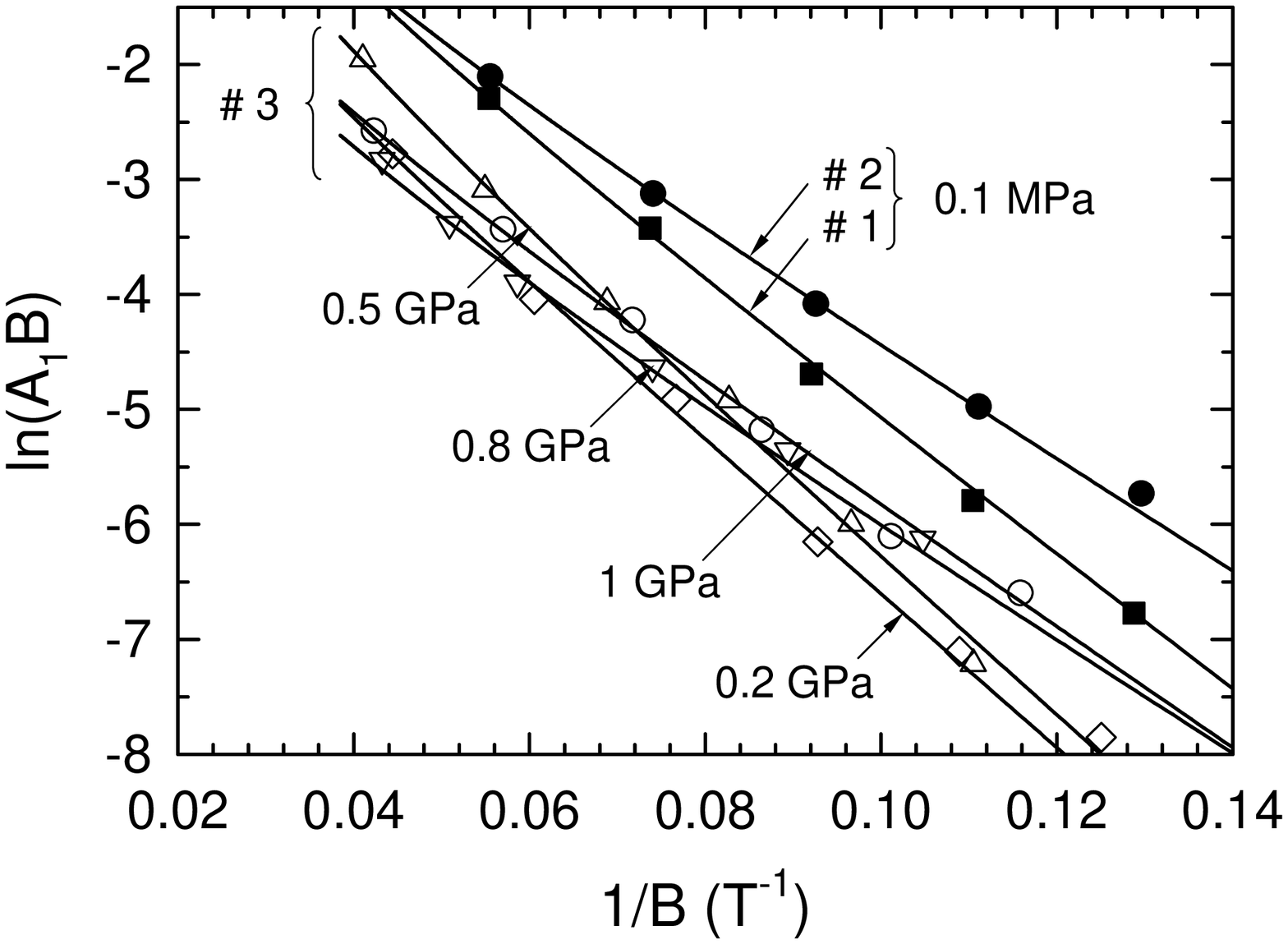}}
\caption{\label{Dingle(P)_F1} Dingle plots of the Fourier
amplitude A$_1$ at 2 K corresponding to the component at the F$_1$
frequency. Solid lines are best fits of Eq. \ref{LK} to the data.}
\end{figure}

Examples of best fits of Equation \ref{approxLK} to the data
collected under applied pressure are displayed in Figure
\ref{fit_m_appliedP}. As it is the case for the ambient pressure
data, a good agreement with Fourier analysis is obtained (see Fig.
\ref{TF_P}). The salient feature of the pressure dependence of the
oscillatory spectra is the progressive decrease of the number of
observed frequencies as the pressure increases. Indeed, only five
and four frequencies can be detected at 0.2 GPa and 0.5 GPa,
respectively. In addition, the slow undulation attributed to F$_0$
cannot be detected above 0.5 GPa. Finally, only three frequencies
labelled F$_1$ to F$_3$ in Figures \ref{fit_m_appliedP}(b) and (c)
are observed at 0.8 GPa and 1 GPa. Both the effective masses (see
Table \ref{table2}) and the Dingle temperatures (see Table
\ref{table2} and Fig. \ref{Dingle(P)_F1}) remain constant within
error bars. The three frequencies observed at high pressure are
connected by the relation F$_1$ + F$_2$ = F$_3$. This point is in
agreement with e.g. Fourier analysis at 1 GPa that yields F$_1$ =
(68 $\pm$ 2) T, F$_2$ = (238 $\pm$ 4) T and F$_3$ = (313 $\pm$ 7)
T, respectively. This suggests that the corresponding orbits are
compensated. As it is the case for the NH$_4$-Fe$\cdot$DMF salt
\cite{Au04}, the high pressure spectra of NH$_4$-Cr$\cdot$DMF can
be accounted for by the band structure calculations assuming the
$\bigodot$ orbit, from which originates the FS, intersects both
along the $(a'^*+b'^*)$ and $a'^*$ directions yielding 3
compensated electron and hole orbits as depicted in Figure
\ref{SF}(c). Within this framework, the frequencies F$_1$, F$_2$
and F$_3$ observed at 0.8 GPa and 1 GPa can be ascribed to the
frequencies labelled F$_a$, F$_{b-a}$ and F$_b$ observed at
ambient pressure in the NH$_4$-Fe$\cdot$DMF salt.

\begin{figure}                                                     
\centering
\resizebox{\columnwidth}{!}{\includegraphics*{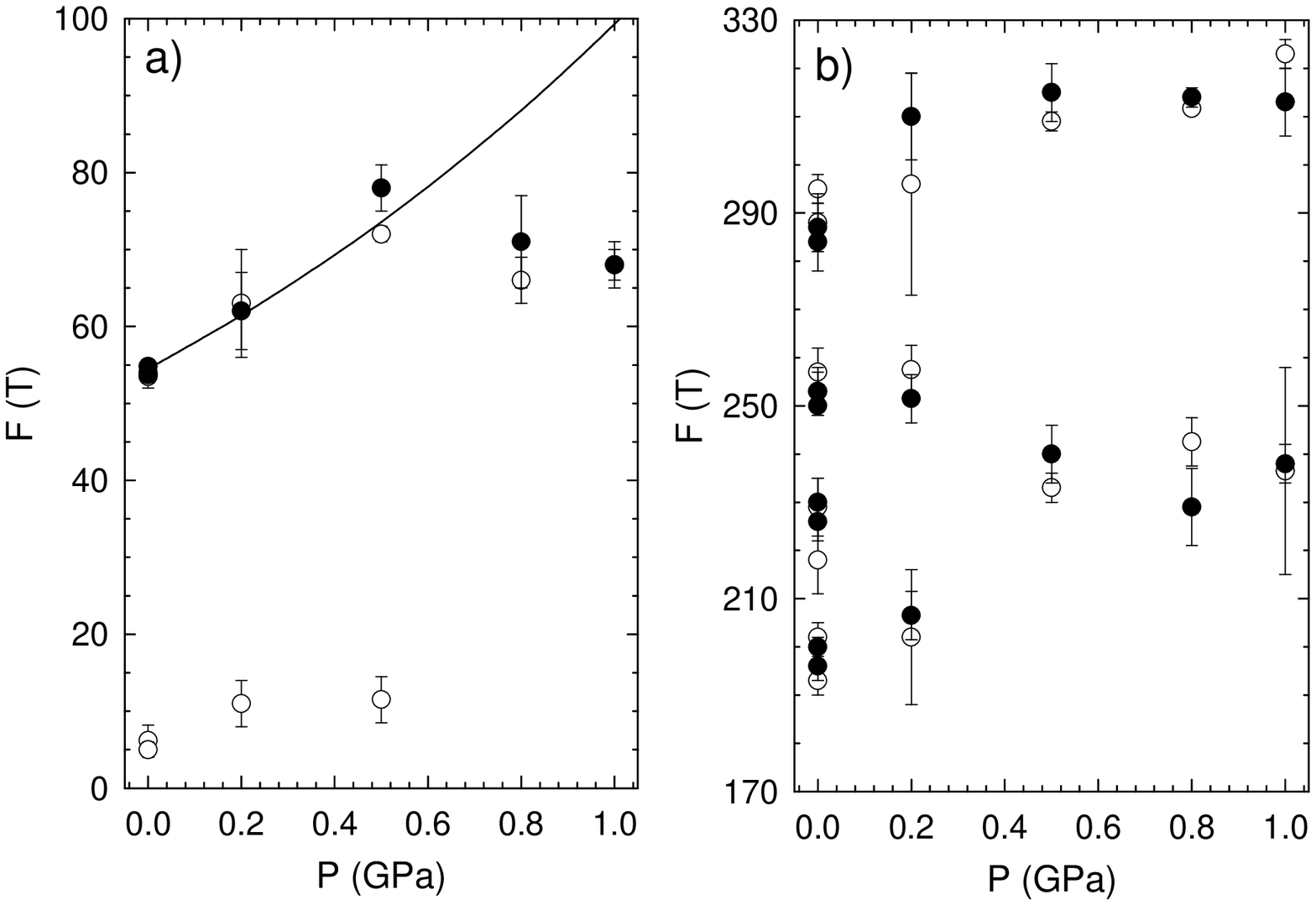}}
\caption{\label{F(P)} Pressure dependence of the frequencies (a)
F$_0$ and F$_1$ and (b) F$_2$ to F$_5$, deduced from Fourier
analysis (solid symbols) and fits of Eq. \ref{approxLK} (open
symbols). The solid line is a fit of the equation F(P) =
F(P=0)exp(-$\kappa$P), with $\kappa$ = 0.6 GPa$^{-1}$, to the data
for F$_1$.}
\end{figure}

The various frequencies observed as the pressure varies are
collected in Figure \ref{F(P)}. Provided a given orbit keeps its
identity as the applied pressure varies, the pressure dependence
of the relevant oscillation frequency can be accounted for by a
relationship of the form d[ln(F)]/dP = $\kappa$ where $\kappa$ is
related to the compressibility tensor. The above relationship
holds for various organic metals based on the BEDT-TTF molecule
with $\kappa$ ranging from 0.14 GPa$^{-1}$ \cite{Br95} to 0.7
GPa$^{-1}$ \cite{Au96}. The solid line in Figure \ref{F(P)}(a) has
been derived assuming this is the case for F$_1$. It is obtained
with $\kappa$ = 0.6 GPa$^{-1}$ which is within the above range. A
clear downward departure from this line is nevertheless observed
above 0.5 GPa. A maximum in the pressure dependence of the
frequency linked to the closed orbit was also reported at $\sim$
0.6 GPa for $\beta$''-(BEDT-TTF)$_2$SF$_5$CH$_2$CF$_2$SO$_3$
\cite{Ha03}. In this latter compound a pressure-induced phase
transition is observed, although at 1.2 GPa. Regarding the other
frequencies, no clear pressure dependence can be derived from data
in figure \ref{F(P)}(b) which certainly accounts for the observed
drastic change of the FS topology as the applied pressure varies.

With regards to frequency combinations linked to either
field-dependent Landau level broadening or chemical potential
oscillations, no such features are observed, at least in the high
pressure range. As a matter of fact, the frequency labelled
F$_{a+b}$, observed in the NH$_4$-Fe$\cdot$DMF salt, which would
correspond to the frequency F$_1$ + F$_3$ in the present case, is
not observed. This can be considered at the light of the large
Dingle temperatures observed (T$_D$ $\sim$ 4 to 9 K under applied
pressure), keeping in mind that oscillations of the chemical
potential are strongly damped by disorder. Since the other source
of frequency combination is the Landau level broadening induced by
coherent magnetic breakdown \cite{LLB}, the absence of such
frequencies suggests that the FS only contains individual orbits,
at least at high pressure.

\section{\label{sec:Conclusion}Summary and conclusion}

The pressure dependence of the FS topology of the
$\beta$''-(BEDT-TTF)$_4$(NH$_4$)[Cr(C$_2$O$_4$)$_3$]$\cdot$DMF
organic metal have been studied up to 1 GPa. The SdH oscillation
spectra observed at ambient pressure are compatible with a FS
composed of up to six individual orbits. A drastic change of the
FS topology is observed under pressure. As a matter of fact, the
number of orbits decreases as the applied pressure increases. At
0.8 GPa and above, only three compensated orbits are observed, as
it is the case for the NH$_4$-Fe$\cdot$DMF salt at ambient
pressure. This feature suggests similar FS in both cases, although
the latter compound exhibits a strongly non-monotonous
temperature-dependent behaviour. This result demonstrates that
such non-monotonous behaviour is not necessarily connected with a
density wave condensation, which was invoked in the case of
H$_3$O-M$\cdot$P \cite{Co04}. At variance with magnetoresistance
data of the NH$_4$-Fe$\cdot$DMF compound and more generally of
many networks of coupled orbits, no frequency combinations, due to
e. g. field-induced chemical potential oscillations, were
observed. This is likely connected to the large amount of disorder
present in the studied crystals as indicated by the very large
measured Dingle temperatures (T$_D$ $\sim$ 7 K).

\begin{acknowledgement}
 This work was supported by the French-Spanish exchange
 programm between CNRS and CSIC (number 16 210) and by Euromagnet under the European Union contract
 R113-CT-2004-506239. We acknowledge Geert Rikken for interesting
 discussions.
 \end{acknowledgement}


%

\end{document}